\documentclass[twocolumn,twoside,slac_two]{revtex4}
\usepackage{graphicx}
\usepackage{fancyhdr}
\begin{document}

\title{A Model of the Spectral Evolution of Pulsar Wind Nebulae}

\author{S. J. Tanaka and F. Takahara}
\affiliation{Department of Earth and Space Science, Graduate School of Science, Osaka University, 1-1 Machikaneyama-cho, Toyonaka, Osaka 560-0043, Japan}

\begin{abstract}
Recent observations suggest that many old pulsar wind nebulae (PWNe) are bright TeV $\gamma$-ray sources without a strong X-ray counterpart.
In this paper, we study the spectral evolution of PWNe taking into account the energy which was injected when they were young for old PWNe.
We model the evolution of the magnetic field and solve for the particle distribution inside a uniformly expanding PWN.
The model is calibrated by fitting the calculated spectrum to the observations of the Crab Nebula at an age of a thousand years.
We find that only a small fraction of the injected energy from the Crab Pulsar goes to the magnetic field, consistent with previous studies.
The spectral evolution model of the Crab Nebula shows that the flux ratio of TeV $\gamma$-rays to X-rays increases with time, which implies that old PWNe are faint at X-rays, but not at TeV $\gamma$-rays.
The increase of this ratio is primarily because the magnetic field decreases with time and is not because the X-ray emitting particles are cooled more rapidly than the TeV $\gamma$-ray emitting particles.
Our spectral evolution model matches the observed rate of the radio flux decrease of the Crab Nebula.
\end{abstract}

\maketitle

\thispagestyle{fancy}


\section{INTRODUCTION}\label{intro}

A pulsar releases its rotational energy as a relativistic magnetized outflow called a pulsar wind.
The pulsar wind collides with surrounding supernova ejecta, forms the termination shock, and creates a PWN \cite{kc84a}.
The acceleration of the pulsar wind particles occurs at the termination shock and the PWN consists of the magnetic field and the ultrarelativistic particles \cite{rg74, kc84b}.
Such a created PWN emits photons ranging from radio to TeV $\gamma$-rays via the synchrotron radiation and the inverse Compton scattering.
Current status of theoretical models as well as observational confrontations is reviewed by Gaensler \& Slane (2006) \cite{gs06}.

The Crab Nebula is one of the best studied PWN at almost all observable wavelengths including its central pulsar, called the Crab Pulsar.
Kennel \& Coroniti (1984) \cite{kc84a} studied the spatial structure of the Crab Nebula, assuming that it is a steady state object (KC model).
They found that the magnetization parameter $\sigma$, the ratio of the electromagnetic energy flux to the particle energy flux just upstream the termination shock of the pulsar wind, must be as small as 0.003 to explain the observed dynamical properties of the Crab Nebula.
Atoyan \& Aharonian (1996) \cite{aa96} succeeded to reconstruct the current observed broadband spectrum of the Crab Nebula by the use of the KC model.

The KC model is not fit to consider the evolution because it is a steady state model.
However, it is very important to consider the spectral evolution of the Crab Nebula.
We need to consider the spectral evolution to explain the flux decrease rate of the Crab Nebula at radio and optical wavelengths \cite{ar85, v07, s03}.
Moreover, we need to understand the spectral evolution of PWNe.
Recent observations found many PWNe which have a variety of the characteristics, such as the ages, the expansion velocities, the morphologies, and the spectra of PWNe.
This variety must be related to the spectral evolution of PWNe.
For example, de Jager \& Djannati-Ata\"i (2008) \cite{dd08} discussed the possibility that the aged PWNe can be seen as the TeV $\gamma$-ray sources without a strong X-ray counterpart.

In this paper, we revisit a spectral evolution model of PWNe.
Although several spectral evolution models of PWNe have been studied \cite{zet08, det08, get09}, several issues still remain to be further clarified.
In our model, the PWN is simply treated as an expanding uniform sphere.
We do not take into account effects of the spatial structure of the PWN, since it is somewhat costly and too detailed to see the whole spectrum.
The energy inside the PWN is injected from the pulsar spin-down energy, which is distributed between the relativistic particles and the magnetic field with a constant ratio.
Our simple model can describe the observed basic features.

We study the spectral evolution of the Crab Nebula for the first application of our model.
The Crab Nebula can be used as a calibrator of our model when it will be applied to other PWNe in future.
In section \ref{model}, we describe our model of the PWN evolution.
In section \ref{crab}, we apply this model to the Crab Nebula.
In section \ref{discussion_conclusion}, discussions and conclusions are made.

\section{THE MODEL}\label{model}
\subsection{Energy Injection}\label{injection}
In this paper, we assume that the PWN is a uniform sphere expanding at a constant velocity $v_{\rm{PWN}}$.
The assumption of a constant velocity $v_{\rm{PWN}}$ is the easiest way to take into account the expansion of the PWN, although the real behavior of the expansion velocity could be more complex as the work made by Gelfand et al. (2009) \cite{get09}.
We consider that the age of the PWN is younger than 10kyr in this paper and that the constant velocity $v_{\rm{PWN}}$ would be a good assumption in this range.

For the components inside the PWN, we assume that the PWN is composed of only magnetic field and relativistic electron-positron plasma, both of which are injected from the pulsar inside the PWN.
We divide the energy injection from the pulsar into the magnetic field energy $\dot{E}_{\rm{B}}$ and the relativistic particle energy $\dot{E}_{\rm{e}}$ using the time independent parameter $\eta$ ($0 \leq \eta \leq 1$).
The fraction parameter $\eta$ is the injection ratio of the magnetic field energy to the spin-down energy.
The fraction parameter $\eta$ in our model is similar to the magnetization parameter $\sigma$ in the KC model, although they are not the same.
The magnetization parameter $\sigma$ is the ratio $\dot{E}_{\rm{B}} / \dot{E}_{\rm{e}}$ at the pulsar wind region immediately upstream the termination shock.
On the other hand, the fraction parameter $\eta$ pertains to $\dot{E}_{\rm{B}} / (\dot{E}_{\rm{B}} + \dot{E}_{\rm{e}})$ into the PWN region uniformly.

For the particle injection, we also need to determine the injection spectrum of the relativistic particles.
Following Venter \& de Jager (2006) \cite{vd06}, we assume that the injection spectrum of the relativistic particles $Q_{\rm inj}(\gamma, t)$ obeys a broken power-law
\begin{equation}\label{eq4}
Q_{\rm inj}(\gamma, t) = \left\{
\begin{array}{ll}
Q_{\rm 0}(t) (\gamma /\gamma_{\rm {b}})^{-p_{\rm 1}} & \mbox{ for $\gamma_{\rm min} \leq \gamma \leq \gamma_{\rm b}$ ,} \\
Q_{\rm 0}(t) (\gamma / \gamma_{\rm {b}})^{-p_{\rm 2}} & \mbox{ for  $\gamma_{\rm b} \leq \gamma \leq \gamma_{\rm max}$ ,}
\end{array} \right.
\end{equation}
where $\gamma$ is the Lorentz factor of the relativistic electron and positron.
We introduce time independent parameters $\gamma_{\rm min}$, $\gamma_{\rm b}$, $\gamma_{\rm max}$, $p_{\rm 1}$ and $p_{\rm 2}$ which are the minimum, break and the maximum Lorentz factors and the power-law indices at the low and high energy ranges of the injection spectra, respectively.
We require that the normalization $Q_{\rm{0}}(t)$ satisfies the following equation
\begin{eqnarray}\label{eq5}
(1-\eta)L(t) = \int_{\gamma_{\rm min}}^{\gamma_{\rm max}} Q_{\rm inj}(\gamma, t) \gamma m_{\rm e} c^{2} d\gamma, 
\end{eqnarray}
where $m_{\rm e}$ and $c$ are the mass of an electron (or positron) and the speed of light, respectively.

\subsection{Evolution of Magnetic Field}\label{mag_evol}
Because the magnetic field lines are stretching and winding, it is difficult to model the evolution of the magnetic field in the context of the uniform PWN.
We have to solve the relativistic magnetohydrodynamics (MHD) equations to determine the realistic magnetic field evolution \cite{det04}.
For simplicity, we assume that the magnetic field evolution are determined in the form of the magnetic energy conservation,
\begin{equation}\label{eq7}
\frac{4\pi}{3} (R_{\rm{PWN}}(t))^{3} \cdot \frac{(B(t))^{2}}{8\pi} = \int_0^{t} \eta L(t') dt'.
\end{equation}

In this model, the magnetic field approximately evolves as $B(t) \propto t^{-1.5}$ for $t > \tau_0$, where $\tau_0$ is the initial spin-down time of the pulsar.
Note that this magnetic field evolution model may be ad hoc, but its behavior is very similar to those adopted in other works \cite{rg74, det09}.
For example, Rees \& Gunn (1974) \cite{rg74} consider the stretching and winding of the magnetic field line inside the uniform PWN and gives $B(t) \propto t^{-1.5}$ for $n = 3$ and $B(t) \propto t^{-5 / 3}$ for $n = 2.5$ for $t > \tau_0$, where $n$ is the braking index of the pulsar.
Another example of de Jager et al. (2009) \cite{det09} gives $B(t) \propto t^{-1.3}$ in their calculation of non-relativistic MHD equations.

\subsection{Evolution of Particle Distribution}\label{part_evol}
We assume that the distribution of the particles in the PWN is isotropic, and then the particle distribution function can be easily volume integrated and is described by $N(\gamma, t)$.
The evolution of the particle distribution $N(\gamma, t)$ is given by the continuity equation in the energy space
\begin{equation}\label{eq12}
\frac{ \partial}{ \partial t} N(\gamma, t) + \frac{ \partial}{ \partial \gamma} \left( \dot{\gamma}(\gamma, t) N(\gamma, t) \right) = Q_{\mathrm{inj}}(\gamma, t). 
\end{equation}

We consider the cooling effects of the relativistic particles $\dot{\gamma}(\gamma, t)$ including the synchrotron radiation, the inverse Compton scattering off the Cosmic Microwave Background Radiation (CMB), and the adiabatic expansion, i.e.,
\begin{equation}\label{eq13}
\dot{\gamma}(\gamma, t) = \dot{\gamma}_{\mathrm{syn}}(\gamma,t) + \dot{\gamma}_{\mathrm{IC}}(\gamma) + \dot{\gamma}_{\mathrm{ad}}(\gamma,t).
\end{equation}
Note that the inverse Compton cooling $\dot{\gamma}_{\mathrm{IC}}(\gamma)$ does not depend on time because we consider that the target photon field is only the CMB.
As the cooling effects, we do not include the inverse Compton scattering off the synchrotron radiation field for simplicity because it never be a more important cooling process than the synchrotron cooling. 
We include it in the calculation of the radiation spectrum.

\section{APPLICATION TO THE CRAB NEBULA}\label{crab}
In this section, we apply our model to the Crab Nebula as the standard calibrator of our model.
The Crab Nebula is one of the best observed PWN at almost all observable wavelengths.

The Crab Pulsar has the period $3.31 \times 10^{-2}\rm s$, its time derivative $4.21 \times 10^{-13} \rm {s} \cdot \rm {s}^{-1}$ and the braking index 2.51.
The progenitor supernova is SN1054, which means the age of the Crab Nebula $t_{\rm age} \approx 950\rm yr$.
Note that $\tau_0 \approx 760 \rm yr$.
Here, we adopt that the Crab Nebula is a sphere of the diameter $\approx 3\rm pc$.
Combining with $t_{\rm age} \approx 950\rm yr$, the constant expansion velocity is $v_{\rm PWN} \approx 1500 \rm km/s$, which is close to the observed expansion velocity of the Crab Nebula.

\subsection{Current Spectrum}\label{crab_current}

\begin{figure}
\includegraphics[scale=0.5]{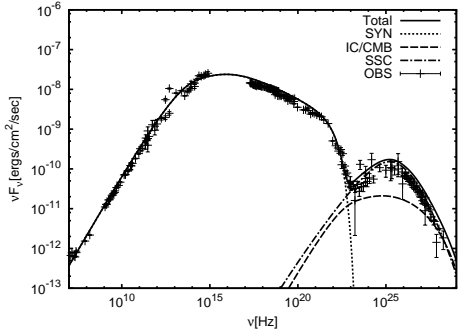}
\caption{Current spectrum of the Crab Nebula in our model and the observational data.
The solid line is the total spectrum which is the sum of the synchrotron (dotted line), IC/CMB (dashed line), and the SSC (dot-dashed line) spectrum, respectively.
Used parameters are tabulated in Table~\ref{tbl-1}. \label{f1}}
\end{figure}

\begin{table}
\begin{center}
\caption{Used parameters to reproduce the current observed spectrum of the Crab Nebula.\label{tbl-1}}
\begin{tabular}{ccc}
\hline\hline
Adopted Parameter & Symbol & Value\\
\hline
Current Period (s) & $\textit{P}$ & $3.31 \times 10^{-2}$\\
Current Period Derivative ($\rm s \cdot \rm s^{-1}$) & $\dot{\textit{P}}$ & $4.21 \times 10^{-13}$\\
Braking Index & $\textit{n}$ & 2.51\\
Age (yr) & $\textit{t}_{\rm age}$ & 950\\
Expansion Velocity (km/s) & $\textit{v}^{}_{\rm PWN}$ & 1500\\
\hline
Fitted Parameter & & \\
\hline
Fraction Parameter $\dot{\textit{E}}_{\rm B} / (\dot{\textit{E}}_{\rm B} + \dot{\textit{E}}_{\rm e})  $& $\eta$ & 0.003\\
Low Energy Index at Injection & $\textit{p}_1$ & 1.5\\
High Energy Index at Injection & $\textit{p}_2$ & 2.45\\
Maximum Energy at Injection & $\gamma_{\rm max}$ & $7.0 \times 10^9$\\
Break Energy at Injection & $\gamma_{\rm b}$ & $6.0 \times 10^5$\\
Minimum Energy at Injection & $\gamma_{\rm min}$ & $1.0 \times 10^2$\\
\hline
\end{tabular}
\end{center}
\end{table}

Figure \ref{f1} shows the current observed spectrum and our calculated one of the Crab Nebula.
The adopted parameters are shown in Table \ref{tbl-1}.
As seen in Figure \ref{f1}, the SSC flux is stronger than the IC/CMB flux at $\gamma$-rays

The fraction parameter $\eta$ governs the absolute values of the fluxes and the flux ratio of the inverse Compton scattering to the synchrotron radiation.
The KC model obtained $\sigma \ll 1$ from the viewpoint of the current dynamical structure of the Crab Nebula, while we determine $\eta \ll 1$ from the viewpoint of the spectral evolution.

In our calculation, the current magnetic field strength of the Crab Nebula turns out to be $B(t_{\rm age}) = 87\mu \rm G$, which is smaller than $B(t_{\rm age}) \approx 300 \mu \rm G$ used by Atoyan \& Aharonian \cite{aa96}.
This difference of the magnetic field strength can be explained as follows.
Atoyan \& Aharonian \cite{aa96} adopted $B(t_{\rm age}) \approx 300 \mu \rm G$ from the KC model and adjusted the particle number to reproduce the observations.
They applied a roughly half a spin-down power compared with the KC model to reproduce the spectrum, thus another half is missing.
On the other hand, all the injected spin-down power is distributed between the particle and the magnetic field energies in our model.

\subsection{Spectral Evolution}\label{crab_evolution}

\begin{figure}
\includegraphics[scale=0.5]{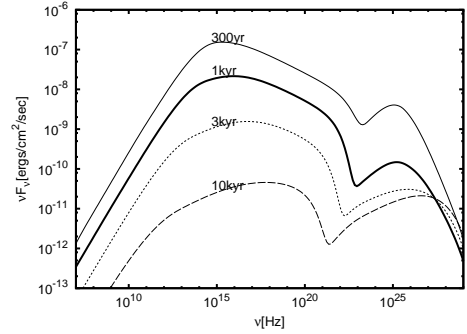}
\caption{Spectral evolution of the Crab Nebula.
The thin solid line is $300\rm yr$ from the birth.
The thick solid, thin dotted and the thin dashed lines are $1\rm kyr$, $3\rm kyr$, and $10\rm kyr$ from birth, respectively.
Each line represents the total spectra which are the sum of the synchrotron, IC/CMB and the SSC spectrum.
Used parameters are the same as in Figure \ref{f1}.
\label{f2}}
\end{figure}

Figure \ref{f2} shows the spectral evolution.
As seen in Figure \ref{f2}, the synchrotron flux decreases with time and the SSC flux also decreases with time owing to decrease of the synchrotron flux, while the IC/CMB flux decreases more slowly than the SSC flux.
This supports the view that old PWNe can be observed as $\gamma$-ray sources which are faint at X-rays.

The radio/optical observations of the Crab Nebula found that the radio/optical flux of the Crab Nebula is decreasing with time.
The inferred rate of the radio flux decrease is $- 0.17 \pm 0.02 \% / \rm yr$ \cite{v07}.
Our model predicts the current rate $\approx - 0.16\% / \rm yr$ and this is almost consistent with the observation.
The inferred rate of the optical continuum flux decrease is $- 0.55 \% / \rm yr$ calibrated at 5000\AA \ \cite{s03}.
Our model predicts the current rate $\approx - 0.22 \% / \rm yr$, this is by a factor of 2.5 smaller than observation.
Note that the trend that the decreasing rate increases with frequency is the same as the observations.

\begin{figure}
\includegraphics[scale=0.5]{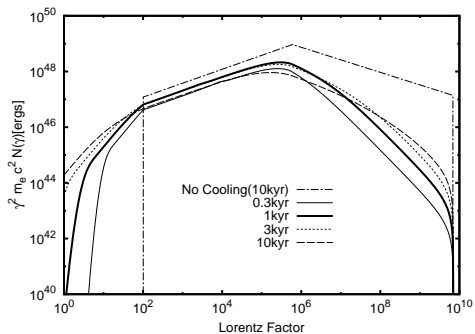}
\caption{Evolution of the particle distribution.
The thin solid line is the distribution at $300\rm yr$ from the birth.
The thick solid, thin dotted and the thin dashed lines are those at $1\rm kyr$, $3\rm kyr$, and $10\rm kyr$, respectively.
The dot-dashed line is the total injected particles until the age $10\rm kyr$.
Used parameters are the same as in Figure \ref{f1}.
\label{f3}}
\end{figure}

To understand the details in Figure \ref{f2}, we show the evolution of the particle energy distribution in Figure \ref{f3}.
For comparison, the total injected particles till 10kyr without the cooling effects is plotted in the dot-dashed line in Figure \ref{f3}. 

In Figure \ref{f3}, we can divide the evolution of the particle distribution in four energy ranges.
First, for $\gamma > 10^8$, the particle number increases with time.
Owing to this increase of the high energy particles, the IC/CMB flux at 10kyr is larger than that at 1kyr above 10 TeV.
Secondly, for $10^5 < \gamma < 10^8$, the evolution of the particle distribution is complex.
The distribution is softer than the injection distribution because of the synchrotron cooling in this range.
Thirdly, for $10^2 < \gamma < 10^5$, the change of the particle distribution is small, the difference between the dashed line (10kyr) and dot-dashed line (10kyr without the cooling effects) is only a factor of two.
This leads to an important conclusion that the radio flux decrease is mainly because of the decrease of the magnetic field.
Combining with the observations of the radio flux decrease, our model of the magnetic field evolution can be near the truth.
Lastly, there exist the particles whose energy is lower than $\gamma_{\rm min}$.
This is because the adiabatic cooling is still effective at low energy.

\section{DISCUSSIONS AND CONCLUSIONS}\label{discussion_conclusion}

\subsection{Discussion}\label{discussion}
Our model of the magnetic field evolution is somewhat ad hoc.
The time dependence of the magnetic field strength $B \propto t^{-1.5}$ for $t > \tau_0$ is assumed.
This is close to other theoretical considerations \cite{rg74, det09}.
Moreover, because our result of the radio flux decrease of the Crab Nebula is almost consistent with the observation, our model of the magnetic field evolution can be near the truth. 

For the injection spectrum of the particle distribution, the acceleration of the particles is an unsolved problem and we adopt the broken power-law injection.
It should be noted that one of the important conclusion in our study that old PWNe can be observed as $\gamma$-ray sources without a strong X-ray counterpart is not affected by the broken power-law assumption.
This is because the low energy particles do not contribute to X-ray and high energy $\gamma$-ray emissions.

\subsection{Conclusions}\label{conclusion}
In this paper, we built a model of the spectral evolution of PWNe and applied this model to the Crab Nebula as a calibrator of our model.
Especially, the magnetic field evolution model is unique and can be close to the reality.

The flux decrease of the $\gamma$-rays is more moderate than radio to X-rays, because the magnetic field decreases rapidly.
This result means that old PWNe can be observed as $\gamma$-ray sources without a strong X-ray counterpart.
De Jager \& Djannati-Ata\"i (2008) \cite{dd08} also suggested that old PWNe can be observed as TeV unidentified sources, because the X-ray emitting particles are cooled more rapidly than TeV $\gamma$-ray emitting particles.
This is not the same reason as our conclusion.

The current observed spectrum of the Crab Nebula is reconstructed when the fraction parameter has a small value $\eta = 0.003$.
This is consistent with the prediction of the magnetization parameter $\sigma \ll 1$ obtained by Kennel \& Coroniti (1984a)\cite{kc84a}.
They obtained $\sigma \ll 1$ from the viewpoint of the current dynamical structure of the Crab Nebula, while we determine $\eta \ll 1$ from the viewpoint of the spectral evolution. 

The smaller value of the current magnetic field $B(t_{\rm age}) = 87 \mu \rm G$ than inferred value $\approx 300 \mu \rm G$ in most of other papers is needed to reconstruct the observed spectrum of the Crab Nebula.
Recent study by  Volpi et al. (2008) \cite{vet08} indicated that the spatially averaged magnetic field strength $B_{\rm mean} \approx 100 \mu \rm G$ in their relativistic MHD simulation.
This is close to our value of the magnetic field strength.

Our model can predict the spectral evolution of the Crab Nebula, the observed flux decrease of the Crab Nebula at radio wavelengths can be explained by our magnetic field evolution model.
The observed flux decrease rate of the Crab Nebula at optical wavelengths is somewhat larger than our model, but the trend that the decreasing rate increases with frequency is the same as observations.

Finally, because we can reproduce the current spectrum of the typical PWN with the spectral evolution model, we can apply this to other PWNe.
We will be able to study whether the observed variations of the PWNe can be understood by the spectral evolution or other effects should be included.

\bigskip 

\begin{acknowledgments}
We are grateful to Y. Ohira for useful discussions.
S. J. T. thanks a The Hayakawa Satio Fund for the support.
This work is partially supported by KAKENHI (F. T. , 20540231).
\end{acknowledgments}

\bigskip 



\begin{thebibliography}{99}
\bibitem{aa96} Atoyan, A. M., \& Aharonian, F. A. 1996. MNRAS, 278, 525
\bibitem{ar85} Aller, H. D., \& Reynolds, S. P. 1985, ApJ, 293, L73
\bibitem{dd08} de Jager, O. C., \& Djannati-Ata\"i, A. 2008, in Nertron Stars and Pulsars: 40 Years After Their Discovery, ed. W. Becker (Berlin: Springer)
\bibitem{det08} de Jager, O. C., Slane, P. O. 2006, \& LaMassa, S. 2008. ApJ, 689, L125
\bibitem{det09} de Jager, O. C., et al. 2009, arXiv:0906.2644
\bibitem{det04} Del Zanna, L., Amato, E., \& Bucciantini, N. 2004. A\&A, 421, 1063
\bibitem{get09} Gelfand, J. D., Slane, P. O., \& Zhang, W. 2009, ApJ, 703, 2051
\bibitem{gs06} Gaensler, B. M., \& Slane, P. O. 2006. ARA\&A, 44, 17
\bibitem{kc84a} Kennel, C. F., \& Coroniti, F. V. 1984a. ApJ, 283, 694
\bibitem{kc84b} Kennel, C. F., \& Coroniti, F. V. 1984b. ApJ, 283, 710
\bibitem{r09} Reynolds, S. P., 2009, ApJ, 703, 662
\bibitem{rg74} Rees, M. J., \& Gunn, J. E. 1974. MNRAS, 167, 1
\bibitem{s03} Smith, N., 2003, MNRAS, 346, 885
\bibitem{vd06} Venter, C., \& de Jager, O. C. 2006, in Proc. 363rd WE-Heraeus Seminar on Nertron Stars and Pulsars, ed. W. Becker \& H.-H. Huang (MPERep. 291)(Garching: MPI extraterr. Phys.), 40
\bibitem{v07} Vinyaikin, E. N., 2007, Astron. Rep., 51, 570
\bibitem{vet08} Volpi, D., Del Zanna, L., Amato, E., \& Bucciantini, N., 2008, A\&A, 485, 337
\bibitem{zet08} Zhang, L., Chen, S. B., \& Fang, J. 2008, ApJ, 676, 1210
\end{thebibliography}
\end{document}